# A biology-inspired model for the electrical response of solid state memristors

24 09 14 arxivphysics

**Agustín Bou\*,[1] Cedric Gonzales,[2] Pablo P. Boix,[3] Antonio Guerrero,[2] Juan Bisquert[3]**

[1] Leibniz-Institute for Solid State and Materials Research Dresden, Helmholtzstraße 20, 01069 Dresden, Germany

[2] Institute of Advanced Materials (INAM), Universitat Jaume I, 12006 Castelló, Spain.

[3] Instituto de Tecnología Química (Universitat Politècnica de València-Agencia Estatal Consejo Superior de Investigaciones Científicas), Av. dels Tarongers, 46022, València, Spain.

Corresponding author: (acatala@uji.es)

**Abstract**

Memristors stand out as promising components in the landscape of memory and computing. Memristors are generally defined by a conductance equation containing a state variable that imparts a memory effect. The current-voltage cycling causes transitions of the conductance, determined by different physical mechanisms such as the formation of conducting filaments in an insulating surrounding. Here we provide a unified description of the set and reset processes, by means of a single voltage activated relaxation time of the memory variable. This approach is based on the Hodgkin-Huxley model that is widely used to describe action potentials dynamics in neurons. We focus on halide perovskite memristors and their intersection with neuroscience-inspired computing. We show that the modelling approach adeptly replicates the experimental traits of both volatile and nonvolatile memristors. Its versatility extends across various device materials and configurations, capturing nuanced behaviors such as scan rate- and upper vertex-dependence. The model also describes well the response to sequences of voltage pulses that cause synaptic potentiation effects. This model serves as a potent tool for comprehending and probing the underlying mechanisms of memristors, by indicating the relaxation properties that control observable response, which opens the way for a detailed physical interpretation.

# 1. Introduction

In the pursuit of advancing computational paradigms, the exploration of novel technologies has become a cornerstone in shaping the future of electronic devices.[1] Among the different computational breakthroughs, resistive random-access memories (ReRAMs), or memristors, have emerged as a promising candidate presenting an unparalleled blend of memory and neuromorphic computing capabilities.[2-6] These memristors exhibit unique characteristics by having both the memory and computing co-located in the same device, storing the information as a modulated resistance. The benefits of this configuration positions them as highly efficient artificially intelligent hardware that can mimic the functions of the human brain.[7-9]

Various emerging technologies have demonstrated this memristive behavior for more complex and demanding in-memory neuromorphic computation schemes.[10, 11] These memristive devices range from metal/oxide/metal structures,[12-14] organic semiconductors,[15-17] complementary metal/oxide/semiconductor (CMOS) compatible silicon-based devices,[18-20] and a wide range of metal halide perovskite formulations.[21-24] Furthermore, depending on the complexity and hardware implementation of the computational schemes and frameworks, specific memristive responses and properties are required. These characteristics vary from nonvolatile binary switching for digital in-memory computing and spiking neural networks,[7-9] to volatile analog switching for brain-inspired computing and artificial neural networks.[25-27] As the field of memristive devices continues to evolve, it is imperative to develop models that provide a deeper understanding of the intricate mechanisms governing their resistive switching behavior.

While the physical mechanisms of the resistive switching and the nature of the internal kinetic processes vary from each material system and configuration, a general formulation for a voltage-controlled memristor is widely accepted, formed by the equations[28-30]

$$I_{tot} = G(u,w)u \qquad (1)$$

$$\tau_{rel}\frac{dw}{dt} = H(u,w) \qquad (2)$$

The equations relate current $I_{tot}$ and voltage $u$ in the device via an internal state variable $w$, and $G(u,w)$ is a general conductance function. Eq. (2) describes the evolution of the variable $w$ in response to change of external stimuli by the relaxation time $\tau_{rel}$ and the general adaptation function $H(u,w)$.

The memristive response of memristors, denoted as *resistive switching*, is defined as the reversible phenomenon of two-terminal elements which changes the resistance upon the application of electrical stimuli.[11, 30, 31] Numerous numerical models ranging from space charge limited current (SCLC),[27, 32] drift-diffusion,[20, 33, 34] and SPICE modelling[35-38] have been proposed to describe the resistive switching of memristors. While being based on the general formulation of Eqs. (1-2), however, the dominant analytical models incorporate a piecewise structure to distinguish the on and off switching cycles, e.g.,[38-42]

$$\tau_+\frac{dw}{dt} = H_+(u,w) \qquad \text{for } u > 0 \qquad (3)$$

$$\tau_- \frac{dw}{dt} = H_-(u, w) \qquad \text{for } u < 0 \tag{4}$$

This distinction is justified since in the set cycle the current must increase, while in the off switching cycle it must decrease. However, the separation in voltage (or current[39]) domains produces some computational troubles, when different types of stimuli must be probed in separate experiments, as is usual in analysis of voltage pulses for synaptical properties. In addition, the piecewise approach obscures the underlying physical properties, which should be continuous.[43]

In lieu of the conventional numerical models for semiconductor devices, the memristive response can be approximated via a biological approach following a neuron model – the Hodgkin Huxley (HH) Model. The HH model describes the different conductance responses due to the intricate dynamics of ion channel currents (leakage current, delayed-rectified potassium ($K^+$) current, and a transient sodium ($Na^+$) current) within neurons responsible for generating the action potentials of neural electrical activity.[44, 45] The HH model incorporates a voltage-controlled activation and deactivation of the channel conductance, but *it uses a single voltage dependent relaxation time* for both activation and deactivation of each memory variable.

The objective of this work is not to develop neurons as in the HH model, but to formulate a general dynamical equation for solid state memristors that can address the response to different types of time dependent stimuli, including pulsed voltage and resistive switching. To this end, we suggest a modelling approach in which each memory variable is controlled by a single relaxation time valid for all domains of voltage and current, as in the HH model.

Firstly, we show the experimental stable resistive switching of a simple methylammonium bromide ($MAPbBr_3$)-based perovskite memristors that will be used to build the model. Further scan rate- and upper vertex voltage-dependencies are shown to establish the representative characteristics of these solid-state memristors. Then, we describe the standard description of voltage-controlled memristors and their corresponding general equations, adopting a continuous relaxation time inspired by HH model. The model validation is facilitated by reproducing the characteristic $I - V$ responses of the representative perovskite-based device, including the scan rate- and upper voltage-dependencies. Moreover, the wide validity of the model is demonstrated by simulating the memristive response of a range of memristor material systems and configurations. We show that the model is able to capture the different device switching properties ranging from threshold volatile memory to bipolar nonvolatile memory.

Finally, in direct test of neuromorphic functionalities, we show that model describes well the response to sequences of voltage pulses that cause synaptic potentiation effects. We conclude that the HH-based memristor model serves as a valuable tool in understanding the underlying mechanisms of a vast range of memristors allowing the direct implementation of these devices in biologically inspired circuits and learning.

## 2. Results

The characteristic $I-V$ curves of the studied perovskite memristor, measured at a scan rate of 1 V s$^{-1}$, in the linear and semi-log scales are shown in Figs. 1a and 1b, respectively, with the inset illustrating the device and measurement configuration. In the linear scale, the device distinctly exhibits a strong inverted hysteresis typically observed in MAPbBr$_3$-based solar cells indicating an inductive time domain response.[46-48] In the semi-log scale, the perovskite memristor features a threshold resistive switching in the positive polarity with an ON/OFF ratio of ~2 orders of magnitude.[10, 49, 50] At low voltages, the memristor is at its initial high resistance state (HRS) also denoted as OFF state. In Figs 1a and 1b the activation process in positive voltages is indicated as (1) in the forward direction and deactivation as (2) in the reverse direction. As the forward voltage scan positive voltages that approaches and passes beyond the threshold voltage of ~ 0.6 V, the device gradually transitions from the HRS to the low resistance state (LRS), also denoted as ON state in binary switching, promoting the SET process.[51] Notably, in the reverse scan direction, the ON state is maintained until the voltage reaches a lower threshold voltage of ~ 0.3 V where the device transitions from the LRS back to the HRS (RESET process). This characteristic resistive switching exhibits volatile memory where the ON state relaxes back to the OFF state upon a sufficient reduction of the applied voltage.[23, 52, 53] The scan rate-dependent characteristic $I-V$ curves (Fig. 1c) further demonstrate the inductive response of the perovskite memristor which exhibits a decreasing forward scan current with increasing scan rate.[54, 55] Moreover, by varying the upper vertex of the $I-V$ measurements (Fig. 1d), the device displays a multilevel/multistate resistive switching suitable for analog volatile memory applications reproducing short-term memory (STM) behavior in neuromorphic systems.[9, 56-59] A broader view of these properties is explained in Ref. [43].

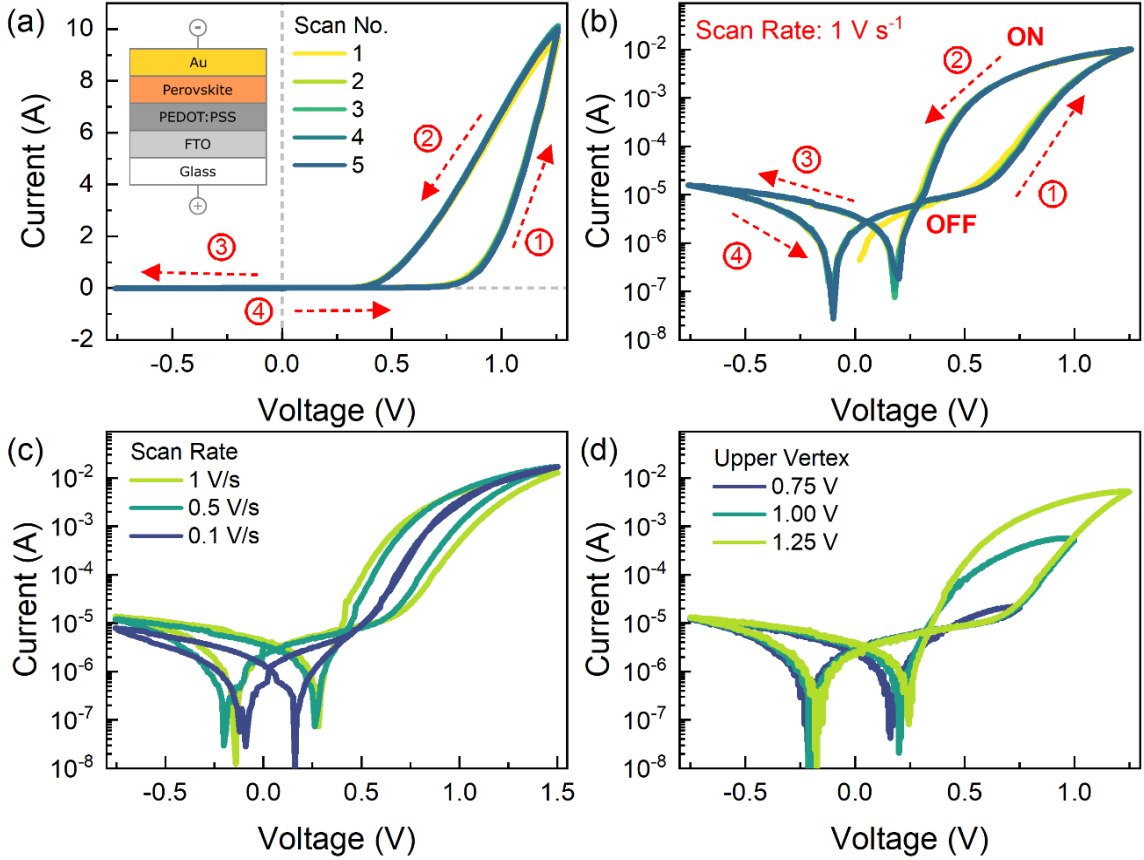

Figure 1. (a) The characteristic $I-V$ response in the linear scale of the FTO / PEDOT:PSS/ MAPbBr$_3$ / Au memristor device with the inset illustrating the schematic diagram of the device configuration, (b) the characteristic $I-V$ response represented in the semi-log scale with the arrows and numbers indicating the scan direction, (c) the scan rate-dependent $I-V$ curves with the reconstructed $I-V$ of the impedance measurements, and (d) the upper vertex-dependent multilevel/analog resistive switching of the memristor device.

## 3. Model.

In order to model the representative characteristic response of memristors, we define the general description of memristive devices.

Based on the general framework of Eqs. (1-2), we now describe a set of characteristics that the model must incorporate. In the memristors used for neuromorphic computation,[60] the system transitions from a low conducting state (of conductance $g_L$) or HRS, to a high conductance state of conductance $g_H$, or LRS, where $g_H \gg g_L$. When the voltage applied is $u$, we have a basic current $I_{tot} = g_L u$, which increases when the resistance switches. We establish the control of conductance by a memory variable $X$ as follows.

$$I_{tot} = [g_L + (g_H - g_L)X]\, u \qquad (5)$$

The equilibrium value of the memory variable $X_{ss}$ is a sigmoidal activation function[40, 61, 62] that corresponds to a Boltzmann open channel probability[63, 64]

$$X_{SS}(u) = \frac{1}{1+e^{-\frac{u-V_t}{V_m}}} + X_0 \tag{6}$$

The threshold voltage $V_t$ defines and translates the state transition of $X_{SS}$ from its initial OFF state $X_{SS} = X_0$ to its ON state $X_{SS} \approx 1$ (Fig. 2a). On the other hand, $V_m$ controls the steepness and voltage window of the state transition where a low $V_m$ exhibits an abrupt state transition at a narrow voltage window, while a high $V_m$ exhibits a gradual state transition at a wider voltage window (Fig. 2b).

We have mentioned the dominant feature of the resistance switching between two conducting states. As demonstrated in the experimental response of the perovskite-based memristor, this threshold resistive switching from the OFF state to the ON state can relax back to the OFF state with the removal or sufficient reduction of the applied electrical stimulus indicating a *volatile memory*.[27] In contrast, another typical resistive switching behavior can maintain both the OFF and ON states for longer durations (up to 10 years) upon the removal of the voltage stimulus indicating a *nonvolatile memory*.[27] These two main types of resistive switching behaviors are taken into consideration in the formulation of our HH-based memristor model as discussed more in detail in the following sections.

The first important point of the model is the need for the high conductance state to be activated at a SET potential, but also to be deactivated at certain RESET potential. Therefore, the $X = 1$ state must decay back to $X = X_0$ to reverse the initial switching. This very basic feature is well acknowledged in solid-state memristor models, yet it is normally achieved by piecewise splitting of the $X$ dynamics into different equations for different voltage domains, as indicated in Eqs. (3-4).[38-41] This approach is not satisfactory as it hides the kinetic origin of the memristor RESET process.

In biological ion models, the central idea is to establish a smooth and continuous model instead of a piecewise one. Therefore, the HH model has provided a better solution. The essential physical property is that the activation time can be modulated by the voltage, apart from the stationary characteristic of Eq. (6). Hence, in the HH model, each relaxation time has a strong voltage dependence. The deactivation of the solid-state memristor at reverse voltage will occur when $u < V_t$, switching back to $X = 0$. However, if $\tau_k$ is long enough, the deactivation is kinetically impeded. As a result, we propose that the kinetic time vanishes at voltage more negative than the threshold RESET voltage $V_{OFF}$. This feature will determine the memristor's volatility at $V_{OFF} \approx V_t$, or if it holds the high conductance below thermodynamic threshold, when $V_{OFF} \ll V_t$.

The second representative characteristics of solid-state memristors is that the kinetical changes may disappear at high voltage if $p$ is already in the ON state. This sets $X = 1$, since $u > V_t$ without additional kinetic changes. Again, here we have $\tau_k \to 0$ after some SET threshold voltage $V_{ON}$ where we observe experimentally a hysteresis free region of the $I - V$ curve at voltages higher than that value. This hysteresis free region above the transition to the high conductive state can be displayed as a linear increase of current with voltage which is discussed further in the later section.

Combining these physical features, we obtain the following model defined by,

$$I = C_m \frac{du}{dt} + [g_L + (g_H - g_L)X]\, u \qquad (7)$$

$$\tau_X(u) \frac{dX}{dt} = X_{ss}(u) - X \qquad (8)$$

where,

$$\tau_X(u) = \frac{\tau_{max}}{1+e^{-\frac{u-V_{OFF}}{V_-}}} - \frac{\tau_{max}}{1+e^{-\frac{u-V_{ON}}{V_+}}} + \tau_{min} \qquad (9)$$

Here, $C_m$ is the device capacitance. The evolution of $X$ is dependent on the kinetic time constant function $\tau_X(u)$. The voltage-dependence of $\tau_X(u)$ is defined by the characteristic maximum kinetic time constant $\tau_{max}$, the minimum kinetic time constant $\tau_{min}$, the threshold RESET voltage $V_{OFF}$ with its corresponding ideality factor $V_-$, and the threshold SET voltage $V_{ON}$ with its corresponding ideality factor $V_+$. The voltage-dependent evolution of the $X_{ss}(u)$ with varying $V_t$ and $V_m$, and $\tau_X(u)$ is illustrated in Fig. 2. The threshold SET and RESET voltages and their corresponding ideality factor parameters affect the $\tau_X(u)$ in the same manner as shown in Fig. 2c. The $V_{OFF}$ defines the threshold voltage where $\tau_X(u)$ transitions from $\tau_X = \tau_{min}$ to $\tau_X \approx \tau_{max}$ with $V_-$ controlling the steepness of transition. On the other hand, $V_{ON}$ defines the threshold voltage where $\tau_X$ reversibly transitions back from $\tau_{max}$ to $\tau_{min}$ with $V_+$ controlling the steepness of transition. Notably, when the threshold voltages are close enough with gradual and wide voltage windows, both transitions can overlap, and the time constant begins to decrease back to $\tau_{min}$ before reaching the $\tau_{max}$.

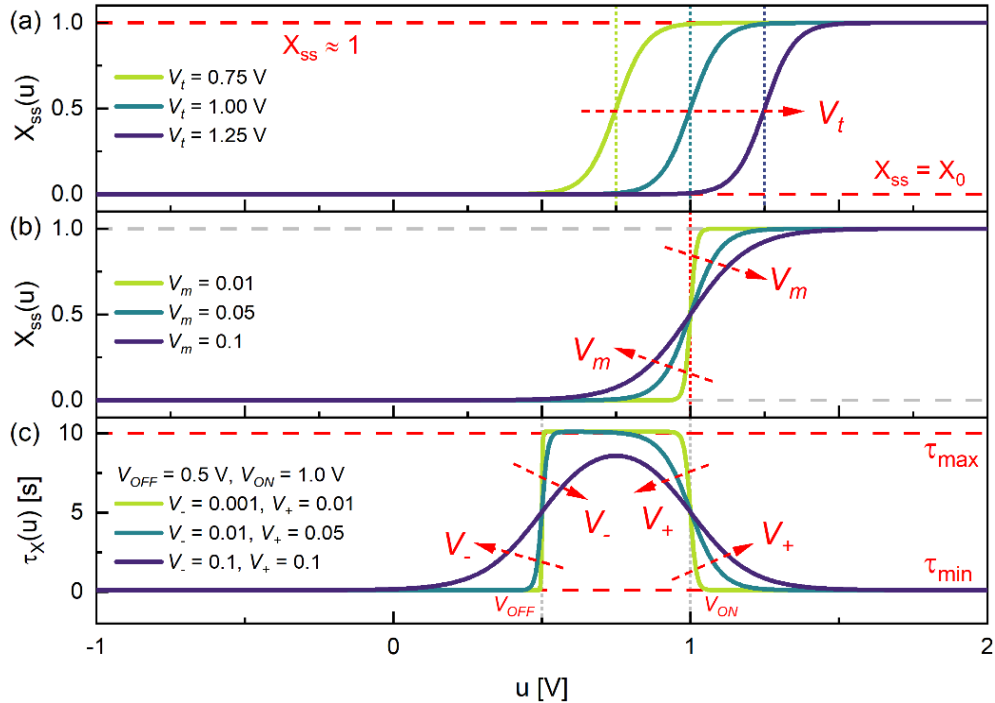

Figure 2. Voltage-dependent internal state variable function $X_{ss}(u)$ of the dynamical neuron-based model with varying (a) voltage parameter $V_t$ and (b) ideality factor parameter $V_m$. Voltage-dependent time constant function $\tau_X(u)$ of at fixed voltage

parameters $V_{OFF}$ = 0.5 V and $V_{ON}$ = 1.0 V with varying ideality factor parameters $V_-$ and $V_+$.

## 4. Discussion

With the theoretical framework explained and the model characteristics illustrated, we implement our biological-inspired model to reproduce the experimental results obtained from the perovskite-based memristor presented in Figure 1. From the characteristic functions of the model, the parameters are selected, as listed in Table 1, to reproduce the experimental trends with the simulated $I-V$ curves shown in Figure 3. Notably, the model is able to replicate the inductive hysteresis in the linear scale (Fig. 3a) and the threshold switching with an ON/OFF ratio of ~2 orders of magnitude at a scan rate of 1 V s$^{-1}$. This volatile threshold switching is obtained by having $V_{ON}$ = 1.05 V and $V_{OFF}$ = 0.75 V both in the positive polarity, while the ON/OFF ratio is controlled by the proper selection of the conductivities $g_L$ and $g_H$. In addition, by using the same parameters and only varying the scan rate $du/dt$, the simulated scan rate-dependent characteristic $I-V$ response of the device is highly consistent with the experimental response exhibiting a decreasing hysteresis with the reduction of scan rate (Fig. 3c).

Lastly, by only changing the upper vertex voltage of the simulated $I-V$ scans, the analog multilevel/multistate resistive switching is obtained in agreement with the actual response of the memristor (Fig. 3d). The corresponding voltage-dependent curves of the pertinent parameters $X_{SS}(u)$, $\tau_X(u)$ and the dynamic evolution of $X(u)$ are shown in Fig. S2. These results demonstrate the model's capacity to reproduce the characteristic $I-V$ response of volatile perovskite-based memristors, including key aspects such as ON-OFF ratio variation, scan rate-dependence, and upper vertex voltage-dependent multistate analog switching.

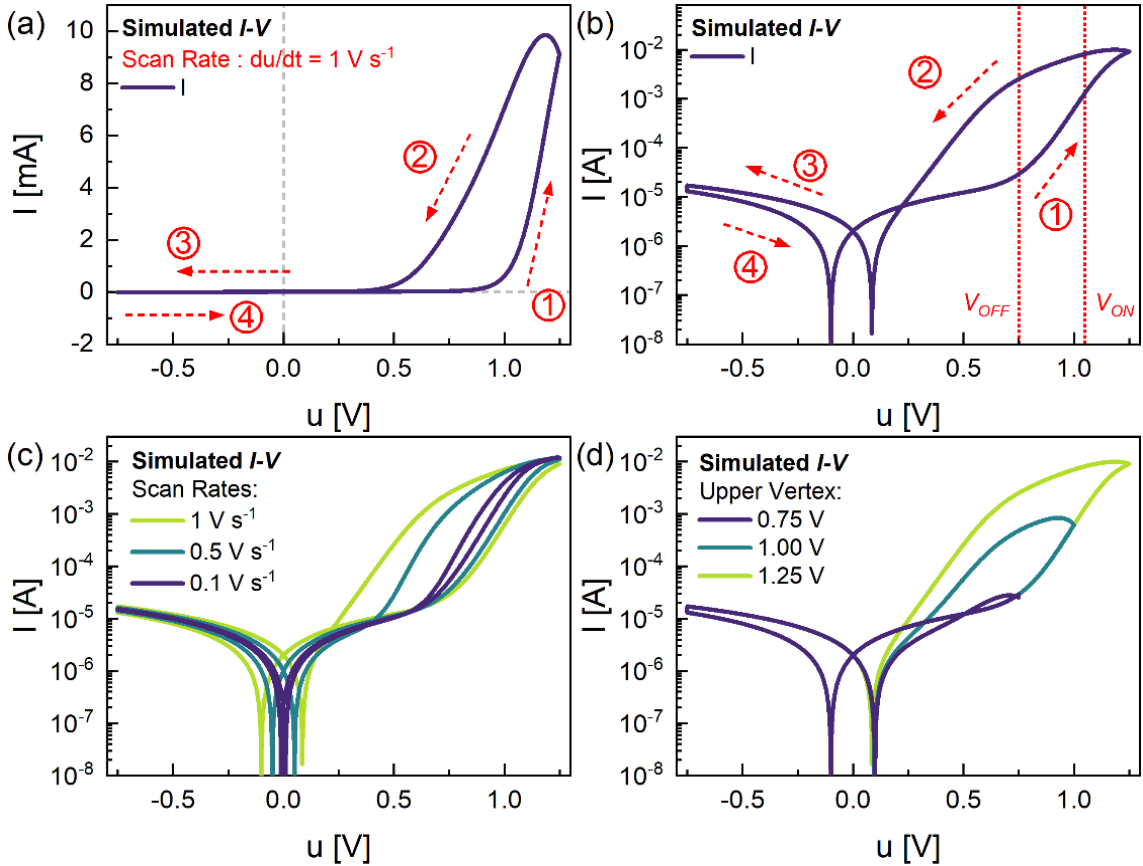

Figure 3. Simulated characteristic $I-V$ response in the (a) linear and (b) semi-log scales with the arrows and numbers indicating the scan direction and sequence, (c) simulated scan rate-dependent $I-V$ response, and (d) simulated upper vertex-dependent multilevel/analog resistive switching using the dynamical neuron-based model.

Table 1. The model parameter list of the simulated characteristic $I-V$ curves in Figs. 3 and 4 that emulate the volatile, nonvolatile gradual, and nonvolatile abrupt resistive switching of perovskite-based memristors.

| Parameters | Volatile (Fig. 3 and Fig. 4a) | Nonvolatile Gradual (Fig. 4b) | Nonvolatile Abrupt 10.1021/acs.jpclett.2c03669 (Fig. 4c) |
|---|---|---|---|
| $C_m$ (μF) | 2 | 10 | 10 |
| $g_L$ (μS) | 20 | 20 | 250 |
| $g_H$ (S) | 0.01 | 0.01 | 0.01 |
| $V_t$ (V) | 1.05 | 0.6 | 0.6 |
| $V_m$ (V) | 0.06 | 0.05 | 0.01 |
| $X_{SS,min}$ | $10^{-6}$ | $10^{-6}$ | $10^{-6}$ |
| $\tau_{max}$ (s) | 0.3 | 10 | 10 |
| $V_{OFF}$ (V) | 0.75 | 0.3 | -0.5 |
| $V_-$ (V) | 0.08 | 0.1 | 0.001 |
| $V_{ON}$ (V) | 1.05 | 0.6 | 0.6 |

| $V_+$ (V) | 0.06 | 0.05 | 0.01 |
| $\tau_{min}$ (s) | 0.08 | 0.1 | 0.1 |

Going a step further, we aim to prove that our model can reproduce the behavior of both volatile and nonvolatile memristors by the precise selection of the characteristic parameters of the kinetic time constant function $\tau_X(u)$. As previously discussed, modifying the value of the threshold RESET voltage $V_{OFF}$ and increasing the time of the LRS to HRS transition can impede the deactivation kinetics. Fig. 4 summarizes three different perovskite memristor configurations exhibiting varying types of memory and their corresponding simulated responses by controlling the threshold voltages $V_{OFF}$ and $V_{ON}$, and their respective ideality factor parameters $V_-$ and $V_+$. The parameter list of the simulated $I-V$ curves are tabulated in Table 1. For the FTO / PEDOT:PSS / MAPbBr$_3$ / Au device exhibiting threshold volatile switching (Fig. 4a), both $V_{OFF}$ and $V_{ON}$ are in the positive polarity with relatively steep but gradual $V_-$ and $V_+$ (Fig. 4d). With these characteristic parameters, $\tau_X(u)$ transitions from $\tau_{min}$ to $\tau_{max}$ and back to $\tau_{min}$ only in the positive voltages (Fig. 4g). Hence, with the removal or sufficient reduction of the applied voltage, the memristor devices relaxes back to the OFF state indicating volatile memory.[10, 65] In contrast, for the FTO / PEDOT:PSS / MAPbI$_3$ / PMMA / Ag / Au exhibiting nonvolatile abrupt binary switching[49] (Fig. 4c), $V_{OFF}$ is negative while $V_{ON}$ is positive with highly abrupt $V_-$ and $V_+$ values (Fig. 4f). These parameters indicate that $\tau_X(u)$ abruptly transitions to $\tau_{max}$ in the negative polarity and again abruptly transitions back to $\tau_{min}$ in the positive polarity. This implies that a negative applied voltage is required to promote the RESET process and the device will stay in the ON state upon the removal of the applied voltage indicating a nonvolatile memory.[10, 65] In between these two extreme cases, an FTO / PEDOT:PSS / MAPbBr$_3$ / Ag / Au device exhibits a nonvolatile but gradual resistive switching (Fig. 4b). For this memristor, both $V_{OFF}$ and $V_{ON}$ are positive similar to the volatile case, however, $V_-$ is highly gradual with a wider transition voltage window (Fig. 4e). The resulting $\tau_X(u)$ begins to transition to $\tau_{max}$ at negative voltages near zero up to positive voltages then transitions back to $\tau_{min}$ without reaching $\tau_{max}$ (Fig. 4h). This suggests that a certain negative voltage is required to fully RESET the device, but the removal of applied voltage would retain the ON state for a specific time defined by the relaxation constant of the kinetic process.

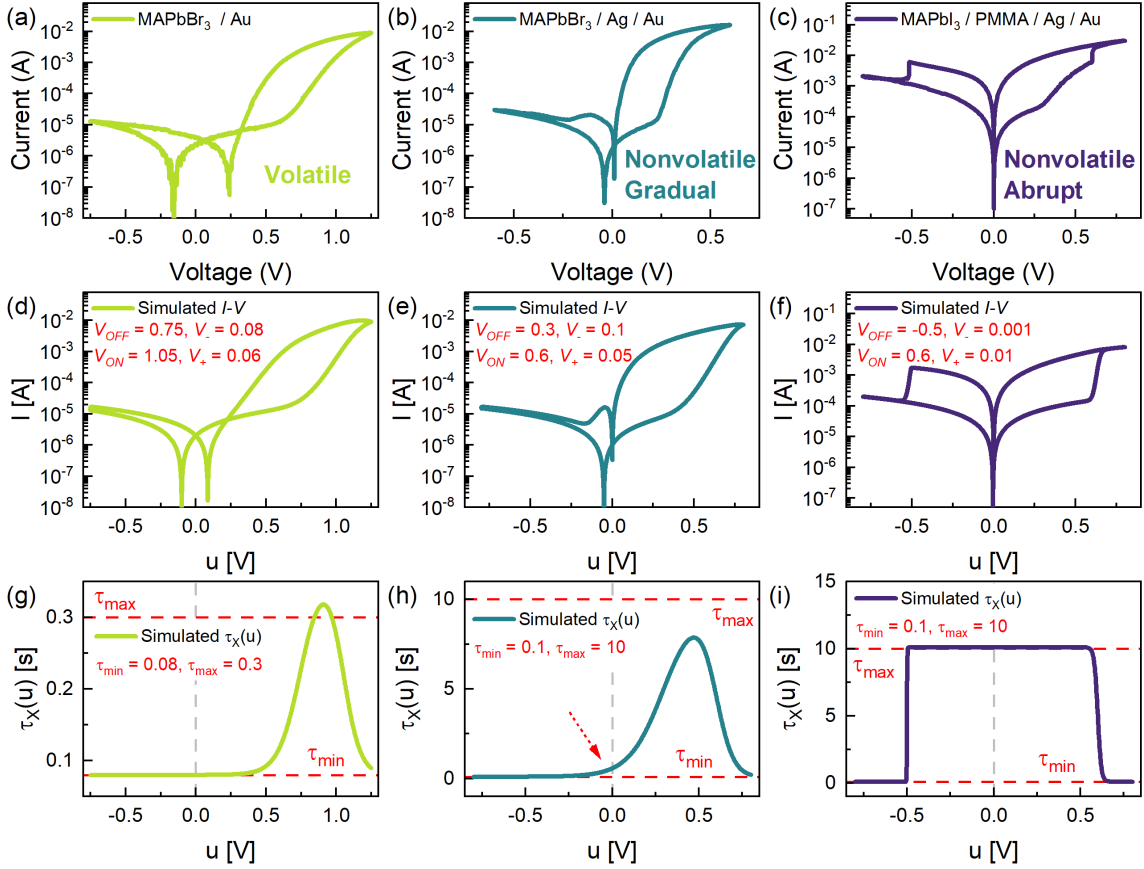

Figure 4. Experimental characteristic $I-V$ response in the semi-log scale of an (a) FTO / PEDOT:PSS / MAPbBr$_3$ / Au memristor exhibiting volatile memory, an (b) FTO / MAPbBr$_3$ / Ag / Au memristor exhibiting nonvolatile memory with gradual SET and RESET, and an (c) FTO / PEDOT:PSS / MAPbI$_3$ / PMMA / Ag / Au memristor[49] exhibiting nonvolatile memory with abrupt SET and RESET. The corresponding simulated $I-V$ responses using the dynamical neuron-based model reproducing the different memory variations - (d) volatile, (e) nonvolatile with gradual SET and RESET, and (f) nonvolatile with abrupt SET and RESET with pertinent parameters indicated.

At this point, we have demonstrated that the model is able to reproduce the volatile, nonvolatile gradual, and nonvolatile abrupt switching of perovskite-based memristors with the proper selection of the characteristic parameters of the kinetic time constant function $\tau_X(u)$. We now extend the implementation of the model to reproduce some of the reported characteristic $I-V$ responses from memristor devices of various material systems and configurations. We show the threshold volatile switching of a high-density fully operational W / SiGe / amorphous-Si / Ag memristor crossbar-array / CMOS system by Kim et.al. that can reliably store complex binary and multilevel 1600 bitmap images (Fig. 5a).[66] By using the biological model, the characteristic $I-V$ curve is reproduced both in the linear and the semi-log scales (Fig. 5c) with the corresponding model parameters listed in Table S2. Most notably, the current beyond the SET voltage features a fully ohmic region which linearly increases with voltage. This ohmic response is due to the device maintaining its high conductance state ($g_L + g_H$) at voltages above this threshold SET voltage as previously discussed.

Another material platform exhibiting volatile threshold switching are diffusive Ag-in-oxide-based memristors by Wang et.al. emulating both the short- and long-term plasticity of biological synapses (Fig. 5b).[3] The different oxide-based memristors emulate the biological $Ca^+$ ion channel dynamics due to the diffusion of Ag metal atoms and spontaneous nanoparticle formation.[3] The biological model is able to simulate the stable resistive switching of the three different oxide materials with of 4 to 10 orders of magnitude ON/OFF ratios (Fig. 5d) with the corresponding model parameters tabulated in Table S2. Valuable insight on the intrinsic properties of the memristor stack can be extracted from the model parameters that would allow direct mechanistic analog with their synaptic functions.

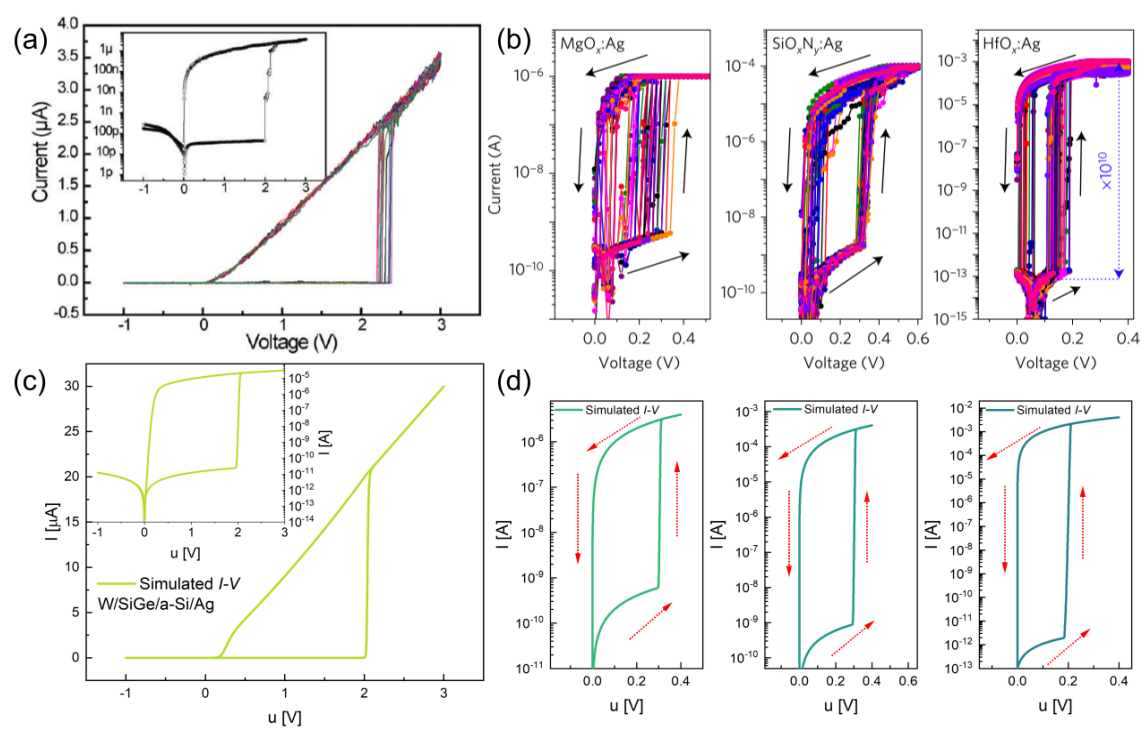

Figure 5. The (a) characteristic $I - V$ response exhibiting a threshold volatile switching of a W/SiGe/a-Si/Ag integrated on top of a CMOS chip with the (c) corresponding simulated curves using the model. Adapted with permission from Kim, K. H., Gaba, S., Wheeler, D., Cruz-Albrecht, J. M., Hussain, T., Srinivasa, N. & Lu, W. A functional hybrid memristor crossbar-array/CMOS system for data storage and neuromorphic applications. *Nano Letters* **12**, 389–395 (2012). Copyright 2012 American Chemical Society. The (b) characteristic $I - V$ curves of various oxide-based memristors with Ag exhibiting threshold volatile memories with the (d) corresponding simulated curves using the model. Reproduced with permission from Wang, Z., Joshi, S., Savel'ev, S. E., Jiang, H., Midya, R., Lin, P., Hu, M., Ge, N., Strachan, J. P., Li, Z., Wu, Q., Barnell, M., Li, G. L., Xin, H. L., Williams, R. S., Xia, Q. & Yang, J. J. Memristors with diffusive dynamics

as synaptic emulators for neuromorphic computing. *Nature Materials* **16**, 101–108 (2017). Copyright 2017 Springer Nature.

The previous two sample responses have been for volatile threshold switching for analog applications in bio-inspired synaptic functions. We next show memristors with bipolar nonvolatile resistive switching for binary in-memory applications. The characteristic $I-V$ response of a p$^{++}$-Si / SiO$_2$ / Ag doped SrTiO$_3$ (ASTO) / Ag memristor device by Ilyas et.al. exhibits a nonvolatile gradual switching (Fig. 6a) and has been demonstrated to mimic simple learning and forgetting behavior in a 7×7 pixel array.[19] This gradual bipolar resistive switching can be simulated by having negative $V_{OFF}$ and positive $V_{ON}$ with relatively higher values of $V_-$ and $V_+$ (Fig. 6c) as tabulated in the parameters list in Table S2. Similarly, a SrRuO$_3$ / Cr-doped SrZrO$_3$ / Au memristor by Beck et.al. has demonstrated reproducible abrupt bipolar switching with long term retention times for nonvolatile 2-bit storage (Fig. 6b).[5] Again, our biology-inspired model is able to reproduce this characteristic response similar to Fig. 6c but with lower values of $V_-$ and $V_+$ (Table S2) resulting to the abrupt SET and RESET processes.

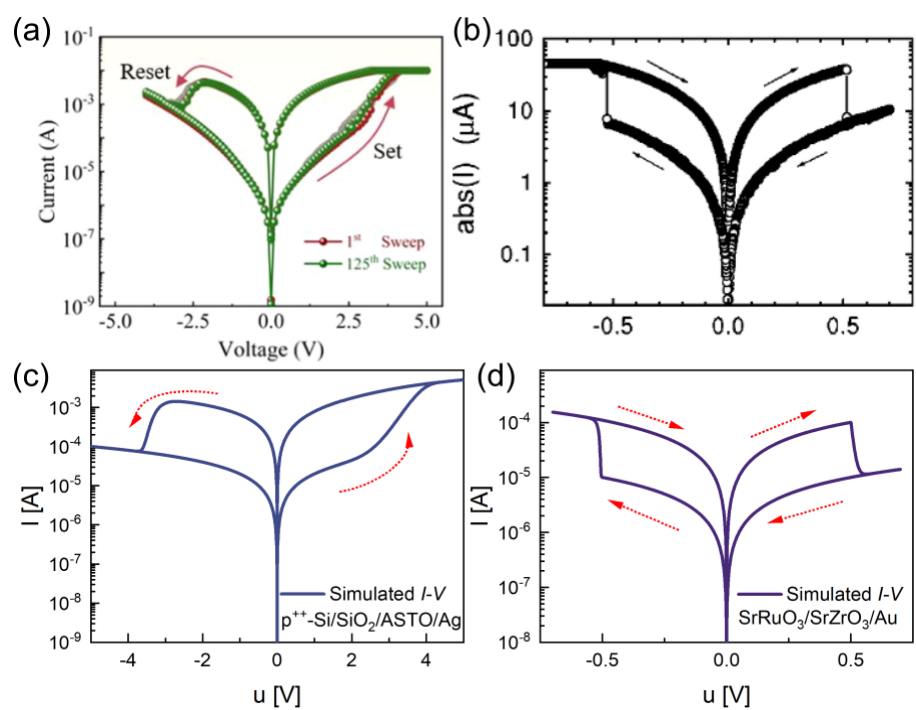

Figure 6. The (a) characteristic $I-V$ response of a p++-Si / SiO$_2$ / ASTO / Ag memristor device exhibiting a bipolar nonvolatile gradual switching with the (c) corresponding simulated $I-V$ response using the model. Adapted with permission from Ilyas, N., Li, C., Wang, J., Jiang, X., Fu, H., Liu, F., Gu, D., Jiang, Y. & Li, W. A Modified SiO$_2$-Based Memristor with Reliable Switching and Multifunctional Synaptic Behaviors. *Journal of Physical Chemistry Letters* **13**, 884–893 (2022). Copyright 2022 American Chemical Society. The (b) characteristic $I-V$ response of a SrRuO$_3$ / Cr-doped SrZrO$_3$ / Au memristor device exhibiting a bipolar nonvolatile abrupt switching with the (d)

corresponding simulated $I - V$ response using the model. Reprinted from Beck, A., Bednorz, J. G., Gerber, C., Rossel, C. & Widmer, D. Reproducible switching effect in thin oxide films for memory applications. *Applied Physics Letters* **77**, 139–141 (2000) with the permission of AIP Publishing.

The wide range of memristive responses that our HH-inspired model has been able to consistently reproduce paves the way for a more holistic and systematic approach in the memristor properties by the analysis of the characteristic $I - V$ curves. As memristors are intended to operate as neuromorphic devices, a biological description of the memristive response based on a smooth and continuous HH model provides a more integrated approach in device characterization and application. It is worth noting that this model in its current form is not able to emulate more complex resistive switching such as unipolar [67, 68] and complementary resistive switching [69, 70] and should not be regarded as a general model for all types of memristors. More complex implementation of the model would require the addition of several conductivities with their corresponding state variables similar to the ionic channels in the HH model. As for its current form, our model provides a simple description of the memristors with only the necessary amount of parameters but is already able to reproduce the most common types of resistive switching. This model will serve as a crucial tool for investigating regular memristors and as a starting point for more complicated memristive behaviors.

## 5. Addressing different types of stimuli: Current potentiation.

The analysis of current-voltage curves at varying scan rates serves as a valuable tool for characterizing the behavior of memristors and provides crucial insights into their switching properties. However, it is essential to recognize that the operational dynamics of memristors significantly differ when employed in practical applications. In real-world scenarios, memristors respond to sequences of voltage pulses with distinct characteristics, undergoing changes in their conductive state based on pulse attributes such as applied voltage or pulse width. With appropriate stimulation, memristors must gradually transition to the ON state, exhibiting an increase in conductivity and current potentiation.

An illustrative example of such potentiation is presented in Figure S2, showcasing the experimental response of a halide perovskite memristor to various voltage pulse trains. The observation reveals that current potentiation does not manifest for low voltages, while it becomes evident at higher voltage levels. Figure 7 further demonstrates the model's capability to replicate these results. Here, we illustrate how the current achieved in each successive voltage pulse surpasses that of the preceding one, indicating a progressive increase in conductivity.

This outcome underscores the model's effectiveness not only in reproducing the current-voltage characteristics of memristors but also in capturing the current potentiation essential for the device's operation in practical applications and synaptic systems.

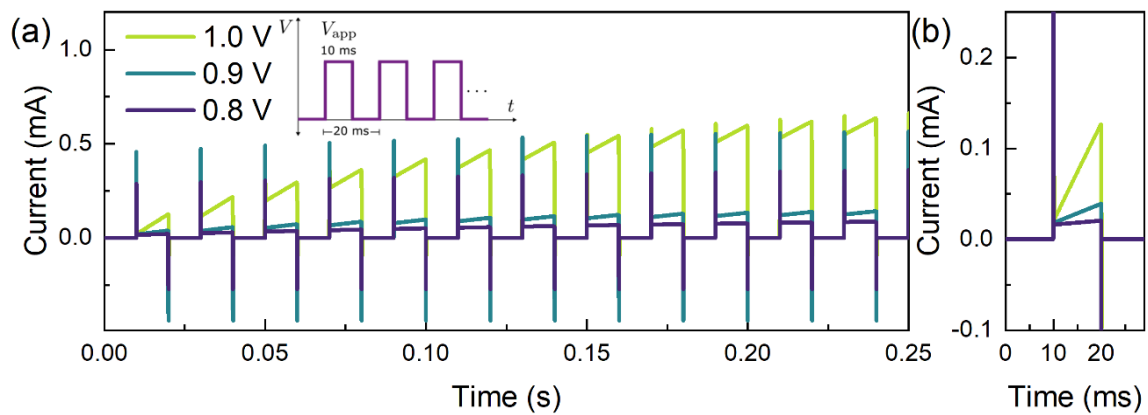

Figure 7. The simulated (a) voltage-dependent transient current response of the FTO / PEDOT:PSS / MAPbBr$_3$ / Au memristor (model parameters in Table 1) using a voltage train stimulus of 10 identical pulses with a pulse width of 10 ms and a pulse period of 20 ms at varying applied voltages potentiation (schematic diagram shown in the inset) exhibiting synaptic, (b) the magnified view of the transient response of the first voltage pulse.

## 6. Conclusions

In conclusion, we have formulated a dynamical model of the characteristic $I - V$ response of memristors, by incorporating internal state functions and variables based on the Hodgkin-Huxley neuron model, emulating the volatile and nonvolatile resistive switching in various device systems and configurations. The model reproduces the experimental characteristics of scan rate-dependent hysteresis and the analog multilevel/multistate resistive switching of a volatile perovskite-based memristor. Not only can the model simulate the various memristive responses in perovskite-based memristors, but it can also be extended in a wide range of memristor material systems and configurations. The development of a biology-inspired model for such memristors, grounded in the HH framework, provides a theoretical foundation that aligns with and successfully captures the observed experimental results. The versatility of the model is showcased through its capability to simulate different types of memristors by systematically adjusting parameters related to the characteristic kinetic time constant function. The presented model serves as a valuable tool for investigating and understanding the underlying mechanisms of memristors in general, and halide perovskite memristors in particular. As the research landscape evolves, this work contributes to the ongoing application and design of memristors, emphasizing their potential impact on the future of electronics and computing.

## Acknowledgments

This work is funded by the European Research Council (ERC) via Advanced Grant 101097688 (PeroSpiker).

# Supporting Information

# A biology-inspired model for the electrical response of solid state memristors

24 09 14 arxivphysics

**Agustín Bou*,[1] Cedric Gonzales,[2] Pablo P. Boix,[3] Antonio Guerrero,[2] Juan Bisquert[3]**

[1] Leibniz-Institute for Solid State and Materials Research Dresden, Helmholtzstraße 20, 01069 Dresden, Germany

[2] Institute of Advanced Materials (INAM), Universitat Jaume I, 12006 Castelló, Spain.

[3] Instituto de Tecnología Química (Universitat Politècnica de València-Agencia Estatal Consejo Superior de Investigaciones Científicas), Av. dels Tarongers, 46022, València, Spain.

Corresponding author: (acatala@uji.es)

## Experimental Details
**Device Fabriction**

The fluorine-doped tin oxide (FTO) substrates (TEC15) were partially etched with zinc powder and a 2 M hydrochloric acid solution. The etched samples were individually brushed to mechanically remove the residues of the etching process. The brushed samples were then subjected to a sequence of 15-minute sonication in deionized water with Hellmanex detergent solution, acetone, and isopropyl alcohol. The cleaned samples were blow-dried using a nitrogen gun.

Prior to the deposition of the poly(3,4-ethylenedioxythiopene) polystyrene sulfonate (PEDOT:PSS), the cleaned samples were subjected to an ultraviolet-ozone (UV-$O_3$) treatment for 15 minutes to further remove organic contamination on the surface and improve the surface wetting. The PEDOT:PSS solution (Clevios P VP. Al 4083) was filtered using a 0.45 µm Nylon filter and was statically spin coated onto the etched FTO substrates for 30 s at 3000 RPM with an acceleration of 1000 RPM/s. The deposited PEDOT:PSS was annealed at 100 °C for 5 minutes then the samples were immediately transported into a nitrogen-controlled glove box in preparation for the $MAPbBr_3$ perovskite deposition.

A 1.4 M $MAPbBr_3$ precursor solution was prepared using $PbBr_2$ (>98%, TCI), and MABr (>99.99%, Greatcell Solar) in 1:4 dimethylsulfoxide (DMSO) (≥99.9%, Sigma Aldrich):N,N-dimethylformamide (DMF) (99.8%, Sigma Aldrich) solution. A 50 µL $MAPbBr_3$ perovskite solution was statically spin coated onto the PEDOT:PSS layer via a two-step antisolvent method: 10 s at 1000 RPM, followed by 40 s at 4000 RPM. A 100 µL toluene (99.8%, Sigma Aldrich) antisolvent was injected 32 s before the spin coating ended. The samples were then annealed at 100 °C for 30 minutes.

For samples with Ag/Au top contact configuration, a 15 nm Ag contact is initially thermally evaporated using a commercial Oerlikon Leybold Univex 250. Finally, an 85 nm gold contact was thermally evaporated as the top contact.

**Voltage-dependent Transient Current Response**

The characteristic $I-V$ curves of the perovskite devices were measured inside a nitrogen-controlled glove box under dark conditions using an Autolab PGSTAT204 with a scan rate of 1 V s$^{-1}$. The voltage-dependent transient current response of the device at an applied volage ($V_{app}$) pulses were measured using 20 voltage pulses with a pulse width ($t_{app}$) of 10 ms.

## Internal Variable and Kinetic Time Constant Functions

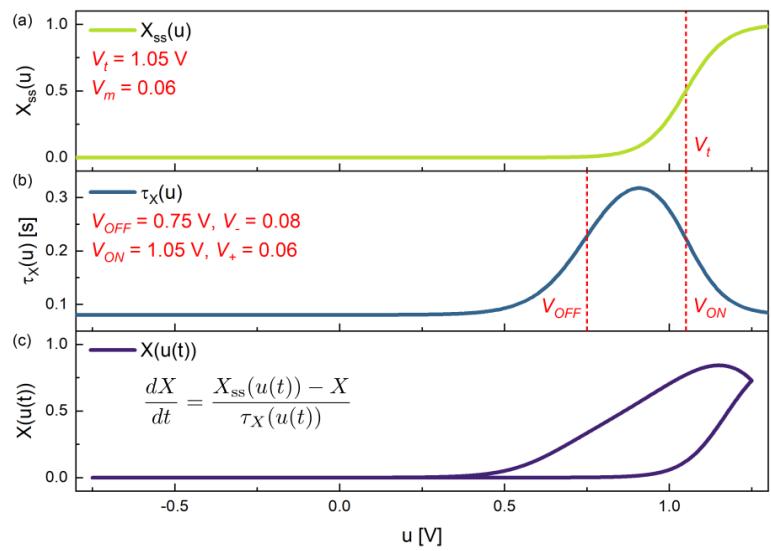

Figure S2. The corresponding (a) internal state variable function $X_{ss}(u)$, (b) the time constant function $\tau_X(u)$, and (c) the time-dependent internal state function $X(u(t))$ of the simulated characteristic $I-V$ response of Fig. 3a with the voltage and ideality factor parameters indicated.

## Model Parameter Lists of Simulated Characteristic I-V Response

Table S1. Parameter list of the simulated characteristic $I-V$ curves in Figs. 5 and 6.

| Parameters | W/SiGe/a-Si/Ag 10.1021/nl203687n (Fig. 5c) | MgOx:Ag 10.1038/nmat4756 (Fig. 5d) | SiOxNy:Ag 10.1038/nmat4756 (Fig. 5d) | HfOx:Ag 10.1038/nmat4756 (Fig. 5d) | p++ Si/SiO2/ASTO/Ag 10.1021/acs.jpclett.1c03912 (Fig. 6c) | SrRuO3/SrZrO3/Au 10.1063/1.126902 (Fig. 6d) |
|---|---|---|---|---|---|---|
| $C_m$ (F) | 1×10$^{12}$ | 1×10$^{12}$ | 1×10$^{12}$ | 1×10$^{12}$ | 1×10$^{9}$ | 1×10$^{9}$ |
| $g_L$ (S) | 2×10$^{-12}$ | 2×10$^{-9}$ | 2×10$^{-9}$ | 2×10$^{-14}$ | 0.00002 | 0.00002 |
| $g_H$ (S) | 0.00001 | 0.00001 | 0.001 | 0.01 | 0.001 | 0.0002 |
| $V_t$ (V) | 2 | 0.3 | 0.3 | 0.2 | 2.75 | 0.5 |
| $V_m$ (V) | 0.005 | 0.001 | 0.001 | 0.001 | 0.3 | 0.001 |

| $X_{SS,min}$ | 0.000001 | 0.000001 | 0.000001 | 1×10$^{-9}$ | 0.000001 | 0.000001 |
|---|---|---|---|---|---|---|
| $\tau_{max}$ (s) | 100 | 100 | 100 | 100 | 100 | 100 |
| $V_{OFF}$ (V) | 0.5 | 0.01 | 0.01 | 0.01 | -2.5 | -0.5 |
| $V_{-}$ (V) | 0.05 | 0.005 | 0.005 | 0.005 | 0.2 | 0.0001 |
| $V_{ON}$ (V) | 2 | 0.3 | 0.3 | 0.2 | 2.75 | 0.5 |
| $V_{+}$ (V) | 0.005 | 0.001 | 0.001 | 0.001 | 0.3 | 0.001 |
| $\tau_{min}$ (s) | 0.1 | 0.01 | 0.01 | 0.001 | 0.1 | 0.1 |

**Experimental Voltage-Dependent Transient Current Response**

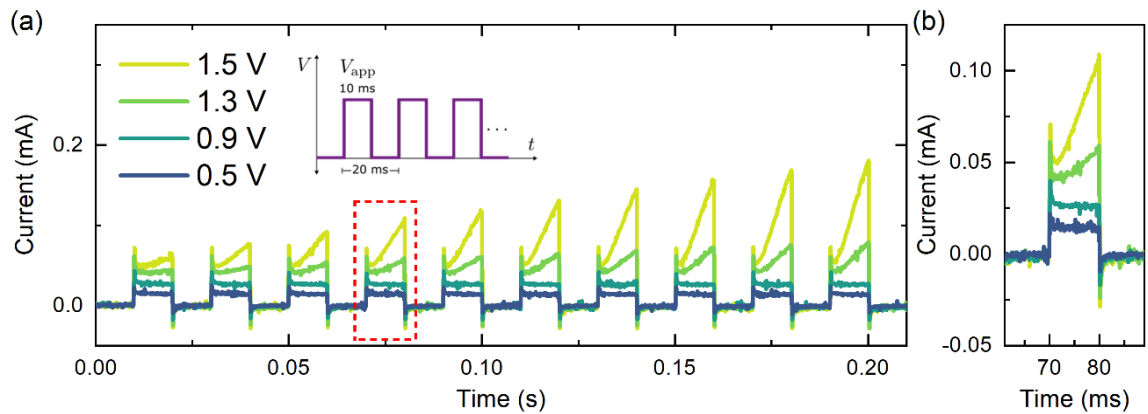

Figure S1. (a) The voltage-dependent transient current response of the FTO / PEDOT:PSS / MAPbBr$_3$ / Au perovskite memristor by applying 10 identical voltage pulses with an amplitude of $V_{app}$ and a pulse width of 10 ms (schematic diagram shown in inset), (b) the magnified view of the transient current response of a single voltage pulse at representative $V_{app}$ levels.